# Creep deformation of WC hardmetals with iron-based binders

Samuel A. Humphry-Baker and Luc. J.M. Vandeperre

*Centre for Advanced Structural Ceramics, Department of Materials, Imperial College London, London SW7 2BP*

ABSTRACT

Iron is a candidate to replace cobalt in WC hardmetals, due to its lower cost and toxicity. A WC-FeCr hardmetal was compression tested at 900-1200 °C. Particular attention is paid to the steady-state creep rates and stress-exponents ($n$) during isostress treatments. Three regimes of stress dependence are observed. Two of these were previously reported for WC-Co: power law creep ($n \approx 3$) at medium stresses; and grain boundary sliding ($n \approx 1$) at higher stresses, generally >100MPa. A previously unreported low stress (<10MPa) regime with an exponent of $n \approx 2$ is also observed. By combining electron microscopy with X-ray diffraction texture measurements, the low stress regime is attributed to viscous flow of the binder, which is accommodated by diffusional creep in the WC skeleton. The mechanism may be applicable to other hardmetals. Compared to analogous WC-Co materials, WC-FeCr shows improved creep resistance below 1000 °C, which can be explained by its lower self-diffusivity, and a lower solubility for WC than Co. However, at temperatures corresponding to liquid eutectic formation (~1140 °C), its creep resistance becomes inferior. These results indicate FeCr may be a suitable replacement for Co provided the eutectic temperature is not exceeded.

## 1. Introduction

Creep resistance underlies the performance of many applications for WC hardmetals. For example, during metal cutting, the tip temperature can exceed 1000 °C [1], at which point WC hardmetals deform extensively by plastic deformation, leading to tool shape change and eventual failure.

Most studies of high temperature deformation have focussed on WC-Co. In these materials deformation is predominantly controlled by the metallic binder below ~800 °C and by deformation and rotation of WC grains above this [2]. A significant increase in fracture toughness also occurs above 800 °C [3]. The dominant creep mechanism depends strongly on the level of applied stress [4–8]. At low stresses, power law creep in the carbide phase dominates, with corresponding stress exponents, $n$, of 3-6 [4,6,7,9]. At higher stresses, cobalt can infiltrate WC/WC boundaries, resulting in grain boundary sliding and a lowering of the stress exponent to $n \approx 1-2$ [4,7]. At higher stresses still, the creep exponent returns a much higher value again, typically $n \approx 6-8$, which has been attributed to a the reactivation of power law creep in the WC skeleton, after all available boundaries for sliding have been exhausted [4].

Creep resistance is maximised at large WC grain sizes and small Co volume fractions. The effect of Co content is most sensitive – for example, doubling the Co volume fraction can result in a 1-2 order of magnitude increase in creep rate [4] – presumably because Co is expected to creep at a rate several orders of magnitude faster than WC [7], although reliable data for the WC phase still does not exist. The grain size dependence is less sensitive, and seemingly has most influence within the boundary sliding regime, where the strain rate obeys an inverse-square relationship with the grain size. Some impurity elements such as Cr and V can significantly improve creep resistance when blended in small concentrations (generally <1 wt.%). This has been related to formation of thin (Cr/V,W)C layers on the surface of WC grains [10]. This surface layer may inhibit deformation by preventing cobalt from infiltrating boundaries.

The choice of metallic binder also affects the creep resistance [11,12]. The ongoing use of Co as binder is problematic because of its extreme price fluctuations and safety concerns over its carcinogenic properties. A safer and more abundant alternative is Fe, however WC-Fe mixtures form a narrow two-phase region over which WC and liquid metal are stable [13]. It is therefore more prone to formation of brittle $M_6C$ or graphite phases when the carbon content is not well controlled. WC-Fe hardmetals appear to have comparable thermophysical properties to WC-Co [14], and according to a recent review [15], the room temperature mechanical properties may be superior. For example, an austenitic binder of 304 stainless steel (Fe-Cr-Ni-Mn) was reported to have an increased hardness compared to Co-bound hardmetals of the same toughness [16].

An open question on the replacement of Co with Fe-based binders is how creep resistance is affected. In FeNi binders, similar mechanisms of deformation have been reported during (hot) turning [17], however there remains a lack of quantitative comparison. Lower creep resistance might be expected due to the lower eutectic temperature in W-C-Fe mixtures compared to W-C-Co [18]. On the other hand, Fe also has a lower solubility for WC, which would have an opposing effect, at least in regimes where diffusion in the binder is a contributing deformation mechanism. Thus, because of multiple creep mechanisms operating simultaneously, extrapolation of properties from those of the constituent phases is not possible. In what follows, a WC-$Fe_{92}Cr_8$ composite is tested under compression in the range 900-1200 °C and its microstructure, grain texture, and flow stress-dependence are assessed as a function of strain rate in order to determine the creep mechanisms. The creep kinetics are compared to conventional WC-Co materials, leading to some recommendations for the improvement of creep resistance in iron-bonded hardmetals.

## 2. Methods

*2.1. Materials*

A plate of $Fe_{92}Cr_8$ hardmetal was provided by Sandvik Hyperion with 90 wt. % WC and 10 wt.% binder. It was sintered for 1 hr at 1400 °C in vacuum. After sintering it had a WC grain size of 0.40 μm and contained ~2% eta phase. Further details on processing and microstructure can be found in a previous report [14]. The largest sides of the plate were parallelised and cut into cuboidal billets of approximately 2.5 x 2.5 x 5mm using electric discharge machining (EDM). The billets were lightly ground using a 1200 grit diamond wheel to remove surface damage. Billet height and cross-sectional area were measured using a digital micrometer with a precision of 1μm.

*2.2. Mechanical tests*

Compression creep tests were performed using a custom-built rig consisting of a 25 kN universal test frame (Instron) and a tungsten-element vacuum furnace (Materials Research Furnaces Inc.). The vacuum was maintained at <5x10$^{-4}$ torr with a Varian Agilent DS102 rotary vacuum pump. The sample was placed between graphite push rods, with SiC spacers either side of it, to prevent any sample-push rod reaction. The spacers were sprayed with hexagonal BN to minimise friction. Samples were heated at 20 °C min$^{-1}$ to the test temperature and held isothermally for 10 min to allow the furnace to reach thermal equilibrium.

Compression tests were performed in two modes: (i) fixed-displacement rate and (ii) fixed-load. For fixed-displacement rate tests, each billet was strained by 4 mm at a fixed rate and then cooled. The test was stopped sooner if the load exceeded 5 kN to avoid failure of the push rods.

For fixed-load tests, billets were first held at a fixed load, generally for 1 h. The load was then increased by a factor of $\sqrt{2}$ and held again. The load was stepped up like this until 1 mm of displacement or 5 kN of load were exceeded. The duration of each hold was initially 1 h however when the strain rate exceeded ~$10^{-4}$ s$^{-1}$ the duration was decreased to limit the plastic strain during each stress increment to 1-2%.

The increase in average sample cross-section was accounted for using the measured plastic displacement under the assumption of conserved volume and transversely isotropic sample expansion with respect to the loading axis.

*2.3. Characterisation*

Crept billets were mounted in resin with flat faces normal to the loading axis. The billets were then ground and polished to a finish of 0.05 µm colloidal silica suspension. Scanning electron microscopy (SEM) was carried out using a Zeiss Sigma 300 instrument operated in Secondary Electron Imaging mode. The grain size was assessed using the arithmetic mean linear intercept method (number average) on at least 150 grains. X-ray diffraction (XRD) patterns were collected on a Panalytical X'Pert powder diffractometer with a Cu radiation source operated at 40kV and 40mA. Patterns were collected in the scan range 10-90° $2\theta$ at a rate of 5° min$^{-1}$.

## 3. Results

*3.1. Fixed displacement rate*

Figure 1 shows fixed displacement rate experiments corresponding to initial strain rates between $\dot{\varepsilon}=10^{-2}$ and $10^{-5}$ s$^{-1}$, collected at $T=1000$, 1100, and 1200 °C. As expected, the flow stress generally increases with increasing $\dot{\varepsilon}$ and decreasing $T$. At low $\dot{\varepsilon}$, the billets deform in a perfectly plastic manner, with approximately constant stress. However, at high $\dot{\varepsilon}$ the billets begin to work harden, with a logarithmically increasing stress profile. The level of $\dot{\varepsilon}$ at which the sample transitions from perfectly plastic to work hardening behaviour decreases with decreasing $T$. For example, at 1200 °C the transition occurs at $\dot{\varepsilon}=10^{-3}$-$10^{-4}$; at 1100 °C it is $\dot{\varepsilon}=10^{-4}$, and at 1000 °C it less than $\dot{\varepsilon}<10^{-5}$. The final engineering strains of the billets are given in Table 1.

**Table 1**
Engineering strain for fixed rate tests at each temperature.

| Nominal $\dot{\varepsilon}$ (s$^{-1}$) | 1000 °C | 1100 °C | 1200 °C |
|---|---|---|---|
| $10^{-2}$ | n.a. | 0.11 | 0.82 |
| $10^{-3}$ | 0.13 | 0.66 | 0.85 |
| $10^{-4}$ | 0.28 | 0.82 | 0.81 |
| $10^{-5}$ | 0.67 | 0.83 | 0.82 |

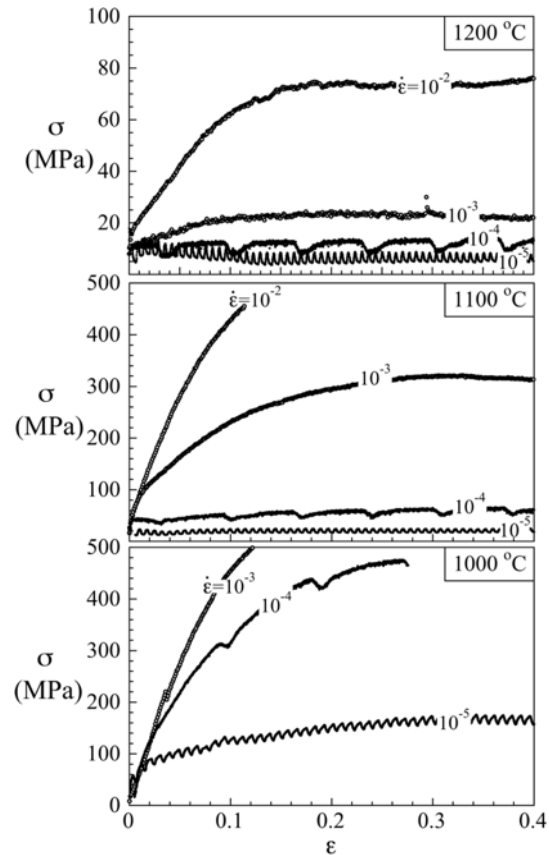

**Fig. 1.** Fixed displacement rate experiments, showing stress, σ, vs strain, ε, at 1000-1200 °C. Nominal strain rate, $\dot{\varepsilon}$ (s$^{-1}$), is indicated on curves. Perfect-plastic flow is seen at low $\dot{\varepsilon}$/high $T$ and work-hardening behaviour at high $\dot{\varepsilon}$/low $T$. Periodic oscillation of the load at the lower strain rates is due to intermittent water-cooling of the instrument.

## 3.2. Fixed load

Figure 2 shows some typical fixed-load experiments at 1100 °C. Part (a) shows curves for 7.1-20 MPa and part (b) for 71-200. Both sets of curves show an initially steeper gradient (primary creep) followed by a minimum in the gradient (secondary creep). At the lower stresses, primary creep lasted a few minutes, while at higher stresses (part (b)), it lasted a few seconds. All subsequently reported strain rate values are taken from linear fits to the secondary creep regime.

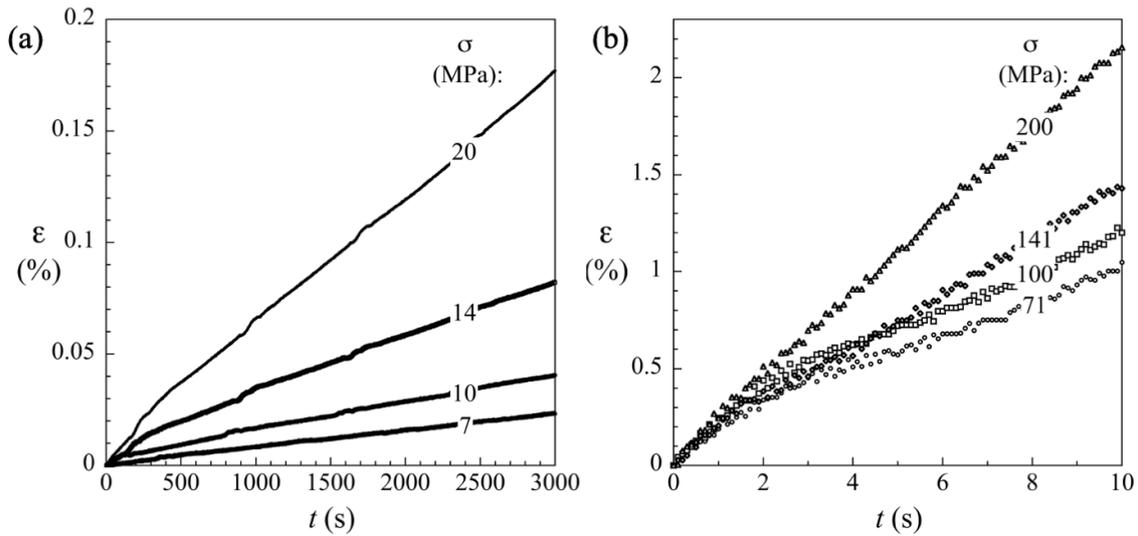

**Fig. 2.** Typical strain data collected at 1100 °C. (a) low stresses are applied for 1h and reach steady state creep after a few minutes. (b) higher stresses applied for ~10s to limit strain to 1-2%, where steady state creep is reached after a few seconds.

Figure 3 shows the steady state creep rates as a function of stress on double-logarithmic scales. The data is fitted with power-law expressions which appear as straight lines. The apparent gradient is the stress exponent of the power-law fit, $n$. The data suggests there are three regimes of $n$: regime I at low $\sigma$, where the $n \approx 2$; regime II at medium $\sigma$, where $n \approx 3$; and regime III at high $\sigma$, where $n \approx 1$. The exact values of $n$ in each stress regime and temperature are given in Table 2. The stresses at which the transition between regimes occurs is temperature dependent. For example, at 1200 °C, the transition from regime I to II is at ~10 MPa, while at 1000 °C it occurs at ~20 MPa. Interestingly, the transition from regime II to III occurs at similar strain rates to the transition from perfectly-plastic to strain-hardening behavior in Fig. 1. For example, at 1100 °C, the transition occurs between $10^{-3}$ and $10^{-4}$ /s.

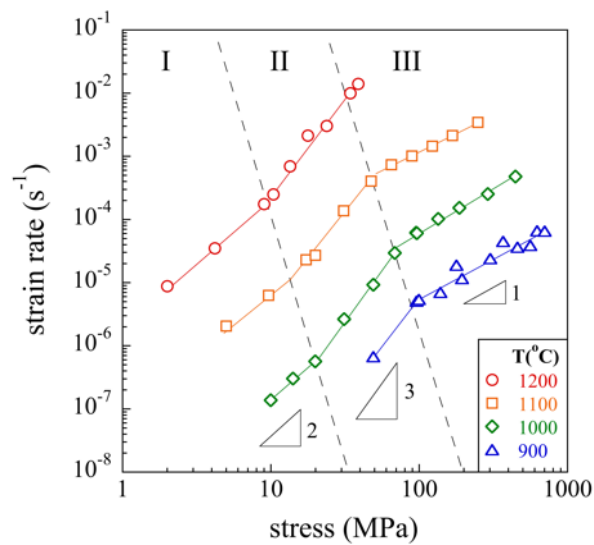

**Fig. 3.** Strain rate as a function of stress for multiple temperatures.

**Table 2**
Numerical value of power law fits to the data in Figure 3, where the regime is denoted by the subscript.

| T (°C) | $n_I$ | $n_{II}$ | $n_{III}$ |
|---|---|---|---|
| 1200 | 2.0 | 2.9 | / |
| 1100 | 1.9 | 3.1 | 1.2 |
| 1000 | 2.1 | 3.0 | 1.3 |
| 900 | / | 3.0 | 1.3 |

To allow calculation of the activation energy for creep in each regime, a general constitutive equation for creep is considered [19]:

$$\dot{\varepsilon} = A \frac{D_{eff} \mu b}{kT} \left(\frac{\sigma}{\mu}\right)^n \tag{1}$$

where $A$ is a dimensionless constant, $\mu$ is the shear modulus, $k$ is the Boltzmann constant, $T$ is the temperature, $\sigma$ is the applied stress and $D_{eff}$ is the effective rate of diffusion

$$D_{eff} = D_0 \cdot exp\left(-\frac{E_A}{RT}\right) \tag{2}$$

with $D_0$ the diffusivity and $E_A$ the activation energy for diffusion, which can vary dependent on the dominant deformation mechanism.

Figure 4 shows an Arrhenius plot for the strain rate in each creep regime. The strain rate has been normalized by all of the terms in equation (1) except for $A$ and $D_{eff}$, thus allowing the effective activation energy to be determined from the gradient of the Arrhenius plot. In order to ensure the only remaining temperature dependence is due to the activation energy, the temperature dependence of the shear modulus was accounted for using data for the WC phase taken from [20], while data for pure Fe from [19] was taken for the binder phase. The modulus of the composite is then calculated using the analytical expression given by Ravichandran [21]. Finally, the creep exponent, $n$, in each regime is taken as the average of the values shown in Table 1. The activation energies of creep deformation in regime (I) and (II) are similar; 550 ± 100 and 580 ± 10 kJ mol[-1] respectively. The value in regime (III) is lower, at only 360 ± 60 kJ mol[-1].

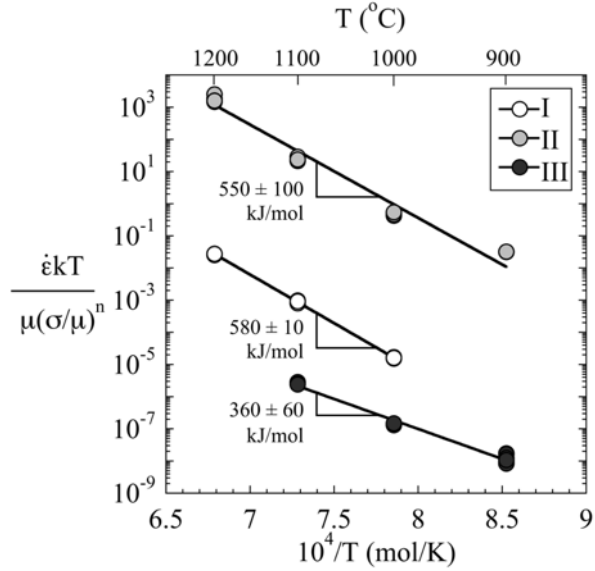

**Fig. 4.** Normalised strain rate as a function of inverse temperature, showing apparent activation energies of 580, 550, and 360 kJ/mol for regimes I, II and III respectively. Uncertainty represents maximum deviation when data at any one temperature is removed. Values of $n$ are taken as the average of the regime-specific values reported in Table 1.

*3.3. Microstructural investigation*

Figure 5 shows SEM micrographs of the billets after the fixed rate tests at 1200 °C. In general, there is an increase in grain size as the strain rates are reduced, from 0.42 μm at $10^{-2}$ s$^{-1}$ to 0.72 μm at $10^{-5}$ s$^{-1}$. Grain growth can be explained by longer test durations (~90 s vs. ~24 h respectively). Grain growth information for all samples tested at 1200 °C is given in Table 3. Looking more closely, the upper row of images shows images taken from the centre of billet, while the lower row shows images taken from the edge. At the highest strain rates (>$10^{-3}$ s$^{-1}$), the images at the edge and centre show no difference, however at lower strain rates the microstructures at the centre are more densified, with a higher volume fraction of WC, while in the edge there is less WC. Thus, FeCr binder appears to have migrated from the centre to the edge of the billet during the test.

**Table 3**
Grain size, $d$, and standard deviation, $s_d$, after fixed displacement rate tests at 1200 °C.

| Nominal $\dot{\varepsilon}$ (s$^{-1}$) | $d$ (μm) | $s_d$ (μm) |
|---|---|---|
| As received | 0.40 | 0.22 |
| $10^{-2}$ | 0.42 | 0.25 |
| $10^{-3}$ | 0.51 | 0.29 |
| $10^{-4}$ | 0.59 | 0.33 |
| $10^{-5}$ | 0.72 | 0.39 |

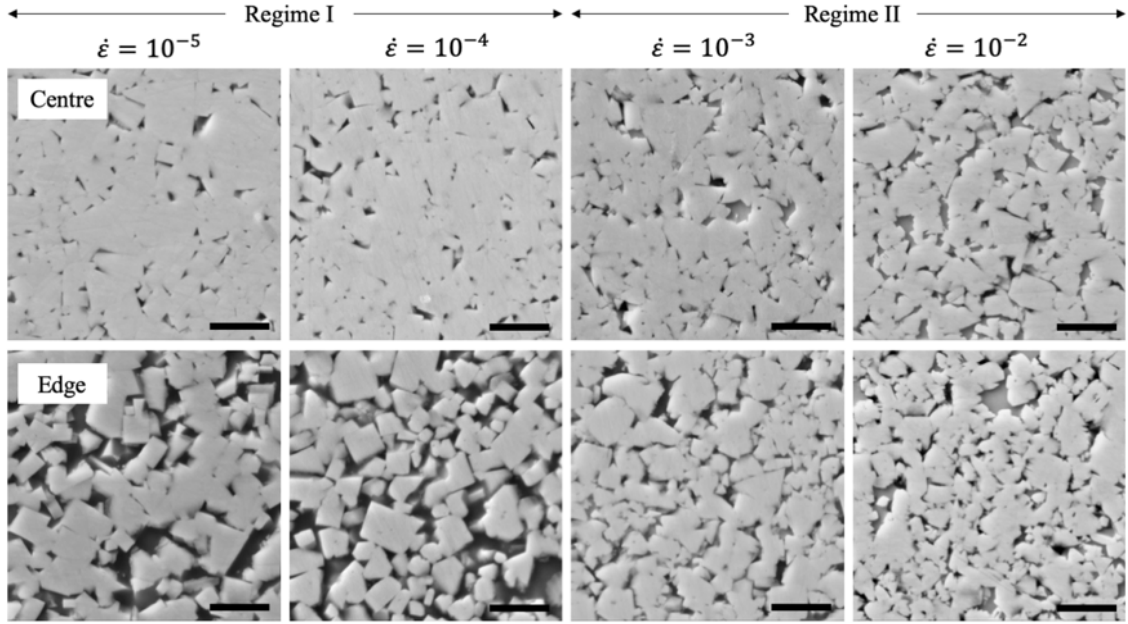

**Fig. 5.** SEM micrographs after engineering strains of ~0.83 at 1200 °C taken from the centre of the billet and the edge (scale bar = 2μm). At the lowest strain rates (regime I), the billet centre becomes lean in binder and the edge is enriched. At higher strain rates (regime II), there is less binder enrichment and at $\dot{\varepsilon}=10^{-2}$ there is no difference. The grain size also increases with decreasing $\dot{\varepsilon}$ (as detailed in Table 3).

Figure 6(a) shows an XRD pattern of the top surface of the sample deformed at 1200 °C and a strain rate of $10^{-3}$ s$^{-1}$ (i.e. within regime II of deformation) and a corresponding pattern of the as-received material. The patterns both contain peaks for WC, α-Fe, and (W,Fe)$_6$C. The relative intensity of the WC peaks has changed significantly. In the as-received state, the WC peak intensities closely match those of the ICDD powder diffraction file. However, in the deformed state the intensity of the basal plane peaks, i.e. (001), (002) and pyramidal plane peak, (101) increases, while the intensity of the prismatic plane peaks, (100) and (110) decreases. Hence the basal and pyramidal planes have aligned preferentially normal to the compression axis, suggesting they are the dominant slip planes during dislocation creep. Such preferential alignment is in agreement with previous measurements on WC-13Co, crept at 1200 °C at a rate of ~$10^{-4}$ s$^{-1}$ [6,22].

Fig. 6(b) plots the ratio of the basal-to-prismatic peak intensities, $I_{basal}/I_{prism}$, after normalisation to the intensity ratios of the as-received material. The average of $I_{(001)}/I_{(100)}$ and $I_{(002)}/I_{(110)}$ has been plotted. For the samples deformed at 1200 °C, there is significant texture development when deformed at $10^{-3}$ s$^{-1}$ or faster (regime II), however when deformed at $10^{-4}$ s$^{-1}$ or slower (regime I) the intensity ratio remains at 1, which indicates that dislocation based slip does not occur in regime I. The samples deformed at 1100 °C show a similar trend although the transition is shifted to lower strain rates. The apparent transition is closer to $10^{-5}$ s$^{-1}$. This is in agreement with the transition between regimes I and II, which according to Fig. 3 also occurs at a deformation rate of $10^{-5}$ s$^{-1}$.

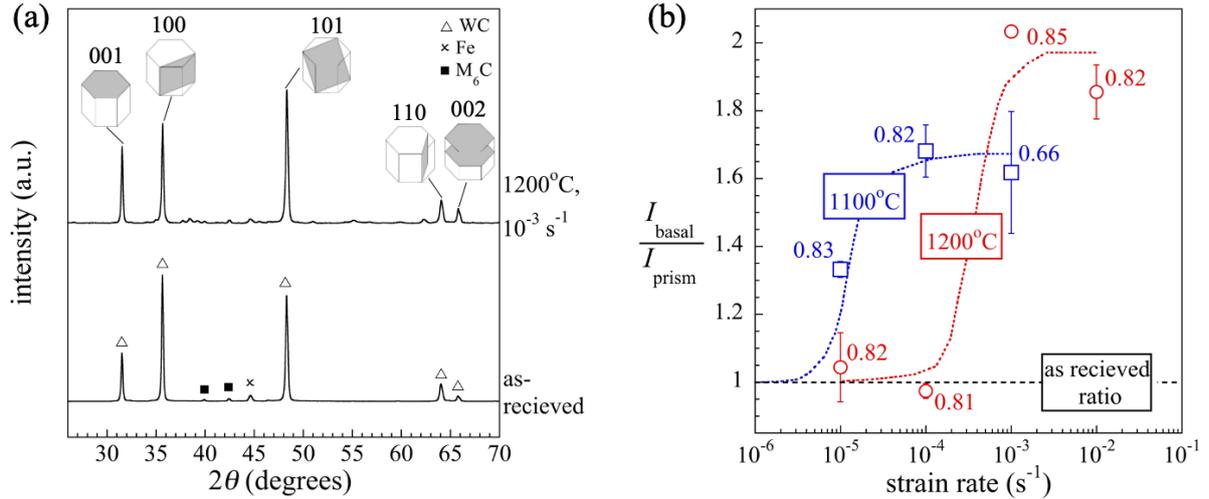

**Fig. 6.** Texture measurements on crept billets: (a) XRD patterns in the as-received state and after engineering strain of 0.85 at 1200 °C and $10^{-3}$ s$^{-1}$; (b) Intensity ratios of the basal and prismatic peaks as a function of strain rate. The average of $I_{(001)}/I_{(100)}$ and $I_{(002)}/I_{(110)}$ have been plotted, error bars representing the deviation between the two. Engineering strains are noted by each datapoint.

## 4. Discussion

The emergence of regime I of the creep exponent (Figure 3) was surprising, because such a low stress regime has not typically been reported for WC-Co [4–6]. The discrepancy could be explained by the minimum applied stresses used in these studies, which were in the range 10-50MPa, while in this study, the minimum applied stress was ~2 MPa. Thus, it is possible such a regime would have emerged, had the authors explored lower stresses. It could also be explained by the lower eutectic temperature in the WC-Fe (1143 °C for WC-10Fe, compared to 1320 °C for WC-10Co [18]). In what follows, we discuss each regime in turn, starting with those more commonly reported, i.e. II and III. The discussion is ended by comparing the creep rates in this study to literature data on WC-Co.

### 4.1. Regime II: power law creep

Regime II is usually associated with power law creep of the WC skeleton [4,6,7,9]. Table 4 lists literature studies where activation energies, $E_{II}$, and creep exponents, $n_{II}$, are reported. The measured $E_{II}$ of 550 kJ/mol is close to that reported for WC-Co alloys of 580 [6], 540-580 [4] and 440-620 kJ/mol [5]. These values are close to the activation energy for self-diffusion in WC. For example, tungsten self-diffusion is reported to be 564 kJ/mol [23] and carbon self-diffusion is calculated from first principles to be 577 and 564 kJ/mol for basal and c-axis diffusion respectively [24].

The measured stress exponent of $n_{II} \approx 3$ is within the range typically observed for WC-Co materials under compression loading, which were reported to be $n_{II} = 2.5$-3.3 by Lee [5] and $n_{II} = 3.3$ Sakuma and Hondo [6]. They are significantly lower than observed during bending experiments, e.g. Lay and Osterstock [4] reported $n_{II} = 4$-6. They are also lower than values reported from tensile experiments by Ueda et al [9] of up to $n_{II} = 5.9$. It is likely that the high exponents measured by Ueda et al [9] are due to the use of tensile loading geometry. It has been established that there is a tension-compression asymmetry in WC-Co and that significantly larger values of $n$ can be expected in tension compared to compression [25]. Similarly, the high exponents measured by Lay and Osterstock [4] can be explained by the use of 3-point bending, since it is known that 3-point bending of ceramics leads to an shift in the neutral axis due to more rapid creep in tension compared to compression, leading to accelerated creep kinetics at higher stresses [26].

**Table 4**
Comparisons of creep constants with literature data on WC-Co hardmetals.

| Binder | Loading method | $E_{II}$ (kJ/mol) | $n_{II}$ | $E_{III}$ (kJ/mol) | $n_{III}$ | Reference |
|---|---|---|---|---|---|---|
| 5-37Co | Bending | 540-580 | 4-6 | 540-580 | 1-2.7 | Lay [4] |
| 13Co | Compression | 580 | 3.3 | 460 | 1.7 | Sakuma [6] |
| 10Co | Tension | 322-439 | 2.9-5.9 | / | / | Ueda [9] |
| 6-20Co | Compression | 440-620 | 2.5-3.3 | 340-480 | 1.4-1.7 | Lee [5] |
| 10FeCr | Compression | 550 | 2.9-3.1 | 360 | 1.2-1.3 | This study |

*4.2. Regime III: grain boundary sliding*

Regime III is usually associated with the onset of grain boundary sliding [4,7]. This is accompanied with a drop in $E$ and $n$. Fig. 4 shows $E_{III}$ =360 kJ/mol, which is reasonably close to that of self-diffusion in gamma iron of 270 kJ/mol [27]. This suggests creep is limited by deformation in the binder phase, which supports the proposed mechanism of grain boundary sliding. It should be noted that our measured activation energy is towards the low end of values usually reported for WC-Co materials. Lay and Osterstock reported no drop in $E_{III}$ [4], while Sakuma and Hondo report a drop from 580 to 460 kJ/mol [6] and Lee from 440-620 to 340-480 kJ/mol [5]. The relatively high values in comparison to this study are somewhat surprising, since the activation energy for Co self-diffusion is 260-273 kJ/mol [28], which is very close to that for Fe of 270 kJ/mol [27]. It is possible that differences arise due to relatively narrow temperature ranges reported in previous studies, e.g. 1100-1200 °C for WC-13Co [6]. Since there is some upwards curvature in the regime III data shown in Fig. 4, when the 900 °C data is discarded the activation energy increases by 60 kJ mol$^{-1}$, therefore it is likely that if a higher temperature range were measured here, a higher activation energy would have been found.

On the other hand, the $n_{III}$ values measured in this study are in excellent agreement with the literature. The most comprehensive study of $n_{III}$ in this stress range was performed by Lay and Osterstock [4], who found that it tended to 1 at very low Co volume fractions, 5 vol. % being the minimum, but increased with increasing Co content, up to about $n_{III} \approx 2.7$ at 37 vol %. The increase in $n$ with Co content was attributed to an increase in the fraction of low-energy boundaries, which are difficult to infiltrate, and therefore do not deform by grain boundary sliding. At the present binder content of 16 vol. %, they reported $n_{III} \approx 1.25$. Thus, our measured value of $n_{III} \approx 1.2$-1.3 at the same binder content is in very good agreement.

*4.3. Regime I: diffusional creep*

With the above mechanisms in mind, we return to regime I, which is not usually reported, and therefore no comparison to the literature on cemented carbides is available. Fig. 4 shows an activation energy of $E_I$ = 580 kJ/mol, which is very similar to that for power-law creep regime, $E_{II}$ = 550 kJ/mol. This suggests that in both regimes, diffusion in the WC skeleton is limiting the overall deformation rate. However, the stress exponent of $n_I \approx 2$ is too low for power law creep, which is usually in range 3-10. Furthermore, as depicted in Fig. 6(b) there is no development of texture in the WC grains, and as shown in Fig. 1, there is no work hardening. Together, this evidence suggests that dislocation-based creep deformation does not occur. It is therefore likely some other mechanism with a lower stress dependence is in operation.

Figure 5 shows that in regime I there was large-scale mass transport of binder to the edge of the specimen when deformed above the eutectic temperature. Such binder migration suggests that a considerable amount of strain is accommodated by viscous binder flow, which is described in Ref. [29]. This mechanism results in a squeezing out of the viscous phase in areas of compressive stress, towards areas of tensile stress. Although the process is considered transitory, i.e. it stops once all of the viscous phase has migrated, the strain at which this occurs is on the order of the total volume fraction of viscous phase [29]. In the case of 10wt.% FeCr, this volume fraction is ~16%. Since the strain at which the transition from region I to II occurred was between 3 and 21%, (the transition strain increased with temperature), it is likely that viscous binder flow occurred in regime I.

The fact that the activation energy for regime I creep was 580 kJ/mol, which is much closer to W self-diffusion in WC (564 kJ/mol [23]) than Fe self-diffusion (270 kJ/mol [27]), suggests that some other process related to the WC skeleton is contributing to the deformation rate as well. Since both phases are fully inter-penetrating, any binder migration would also require WC deformation. As discussed, Figs. 1 and 6 suggest dislocation motion is not occurring, therefore it is likely that any WC deformation is accomplished by diffusion. The solution-precipitation mechanism has been proposed to occur in ceramics with a glassy second phase [30]. The mechanism is similar to that shown to occur during the final stage of liquid phase sintering [31]. We propose that a similar mechanism could be in operation here, and that its dominance will increase as the binder viscosity increases (temperature decreases). The fact that the activation energy is close to that for WC diffusion suggests that the process is limited by interface reaction, rather than diffusional transport through the binder.

It is possible that diffusional creep is relevant to Co-based hardmetals too. The mechanism was proposed to explain anomalous strengthening effect of high Co concentrations in WC-Co when tensile tested at very high temperature (1200 °C) [32]. Interestingly, the applied strain rates were $7 \times 10^{-5}$ s$^{-1}$, which is comparable to the observed strain rates in regime I in this study at 1200 °C, which were on the order of $10^{-4}$-$10^{-5}$ s$^{-1}$. It is therefore possible that this mechanism is relevant to a variety of metallic binder systems.

*4.4. Comparison to WC-Co*

We now compare the creep rates observed in this study to those reported for WC-Co materials of a similar grain size, $d$, and binder weight fraction, $f_{Co}$. High temperature data between 1100 and 1200 °C is taken from Lee [5] and Sakuma and Hondo [6], for materials with $f_{Co}$ = 13 and grain sizes of $d$=0.8 μm and $d$=1.3 μm respectively. Low temperature data between 900 and 1000 °C is taken from Smith and Wood [8], based on materials of $f_{Co}$ = 12 and at grain sizes of $d$=2.2, 3.3 and 4.5 μm.

Figure 7 shows a comparison of the creep rates reported in WC-Co with the present study. The literature data (closed symbols) between 1100 and 1200 °C, plotted in the upper boxes, show that the Co-bonded materials outperform WC-FeCr at the highest temperatures. The degree to which WC-Co outperforms WC-FeCr increases with increasing temperature. The WC-Co materials have slightly higher binder content ($f_{FeCr}$=10, vs. $f_{Co}$=13), which would increase creep rates: Lay and Osterstock [4] found $f_{Co}$=14 materials crept 3-4 times faster than $f_{Co}$=10 materials at 91 MPa. However, the Co-based materials also have a larger grain size (0.8-1.3 μm, vs. 0.4 μm in the present study), which would have the opposing effect. The same study reported a $d_{WC}^{-2}$ dependence, which corresponds to 4-10 times slower creep rates in the coarser-grained materials. Thus, the comparison in Fig. 7 perhaps slightly underestimates the performance of WC-FeCr. However, the effect will be small, therefore it is likely WC-Co still performs better at 1200 °C. The reason for this is most likely the lower eutectic temperature in iron-bonded hardmetals – 1143 °C for WC-10Fe, compared to 1320 °C for WC-10Co [18], meaning that at 1200 °C, the WC-FeCr will contain a significant amount of molten binder phase, which would dramatically increase the rates of grain boundary sliding (regime III) and of viscous binder flow (regime I).

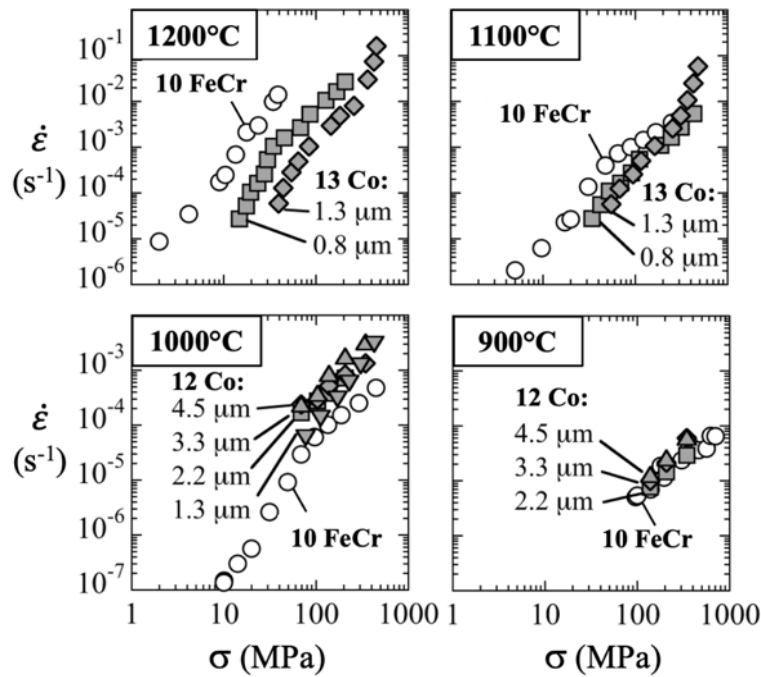

**Fig. 7.** Comparison of creep rates of WC-13Co and WC-10FeCr. Data taken from Smith and Wood [8], Sakuma and Hondo [6] and Lee [5].

As the deformation temperature decreases from 1000 to 900 °C, shown in the lower plots, it is clear that the relative performance of WC-FeCr begins to improve. This is particularly pronounced at 1000 °C, where creep rates are a factor of 10 or so lower. The improved performance of WC-FeCr is likely more pronounced than Fig. 7 suggests, given the much finer grain size studied here, which is 2-10 times smaller than the Co-based materials in Ref. [8]. The lower creep rates in WC-FeCr at these temperatures could be due the lower solubility for WC in Fe compared to Co at high temperature [33]. It could also be due to the presence of Cr, which is known to segregate to WC grain boundaries in WC-Co and form a cubic (Cr,W)C surface layer which inhibits grain growth and creep deformation by preventing infiltration of WC/WC boundaries [34]. While such Cr segregation is yet to be systematically studied for WC-FeCr composites, in a previous work on this material [35], we found 100nm Cr-C precipitates dispersed throughout the binder, therefore it is possible that these nanoscale precipitates strengthen the material at more moderate temperatures.

## 5. Conclusions

We have characterised the creep performance of WC-FeCr, which is a candidate replacement for WC-Co in high temperature applications.

Under the range of temperatures (900-1200 °C) and stresses (~5-500 MPa) studied, the material follows similar regimes of stress dependence observed in WC-Co: grain boundary sliding at the highest stresses (regime III) and power-law creep at more moderate stresses (regime II).

At the very lowest stresses studied, there is the emergence of a lower stress exponent (regime I), which is not yet reported. In this regime there was considerable long-range migration of the metallic binder and no texture development (unlike regimes II and III), suggesting the mechanism is not dislocation-based and involved extensive binder transport. It is concluded that viscous flow of the binder is a contributing deformation mechanism, with the overall deformation rate being limited by solution-reprecipitation of WC.

The relative performance of WC-FeCr and WC-Co depends strongly on temperature. At high temperatures, WC-FeCr shows inferior creep resistance, which is attributed to its comparatively low liquid eutectic temperature (1143 °C for WC-10Fe, compared to 1320 °C for WC-10Co). At moderate temperatures, WC-FeCr appears to outperform WC-Co, which can be explained by the combination of lower WC solubility in Fe compared to Co, and the presence of Cr, which is a known creep suppressant.

More work is needed to understand the role of Cr in creep deformation of hardmetals with Fe-based binders, possibly involving characterisation of the distribution of Cr before and after hot deformation. Characterisation of the creep performance of other Fe-based binders such as FeNi may be useful, particularly as the eutectic temperature of WC-FeNi increases with Ni content.

## Data availability

The raw data required to reproduce these findings will be made available on request.

## Acknowledgments

The authors wish to thank Jessica Marshall and Jonathan Fair of Sandvik Hyperion for providing the WC-FeCr material. S. A. Humphry-Baker was financially supported by the Imperial College Research Fellowship and by EPSRC funding through the program grant (EP/K008749/1) Materials Systems for Extreme Environments (XMat).